\documentclass{PoS}

\title{Performance of the SST-1M telescope for the Cherenkov Telescope
  Array observatory}

\ShortTitle{SST-1M telescope performance}

\author{R.~Moderski$^c$,W.~Bilnik$^k$, J.~B{\l}ocki$^g$,
  L.~Bogacz$^e$, T.~Bulik$^d$, F.~Cadoux$^a$, A.~Christov$^a$,
  M.~Chru{\'s}li{\'n}ska$^d$, 
  M.~Cury{\l}o$^g$, D.~della~Volpe$^a$, M.~Dyrda$^g$, Y.~Favre$^a$,
  A.~Frankowski$^c$, {\L}.~Grudnik$^g$, M.~Grudzi{\'n}ska$^d$,
  M.~Heller$^a$, B.~Id{\'z}kowski$^b$, M.~Jamrozy$^b$, M.~Janiak$^c$,
  J.~Kasperek$^k$, K.~Lalik$^k$, E.~Lyard$^f$, E.~Mach$^g$,
  D.~Mandat$^m$, A.~Marsza{\l}ek$^{h,b}$, J.~Micha{\l}owski$^g$,
  T.~Montaruli$^a$, A.~Neronov$^f$, J.~Niemiec$^g$, M.~Ostrowski$^b$,
  P.~Pa{\'s}ko$^h$, M.~Pech$^m$, \speaker{A.~Porcelli}$^a$,
  E.~Prandini$^f$, E.~Pueschel$^n$, P.~Rajda$^k$, M.~Rameez$^a$,
  P.~Rozwadowski$^d$, 
  E.~jr~Schioppa$^a$,
  P.~Schovanek$^m$, 
  K.~Seweryn$^h$, K.~Skowron$^g$, V.~Sliusar$^i$, M.~Sowi{\'n}ski$^g$,
  {\L}.~Stawarz$^b$, M.~Stodulska$^b$, M.~Stodulski$^g$,
  S.~Toscano$^{f,l}$, I.~Troyano Pujadas$^a$, R.~Walter$^f$, M.~Wi{\c
    e}cek$^k$, A.~Zagda{\'n}ski$^b$, K.~Zi{\c e}tara$^b$,
  P.~{\.Z}ychowski$^g$ for the CTA Consortium\thanks{Full consortium
    author list at http://cta-observatory.org}\\
\llap{$^a$}DPNC -- Universit{\'e} de Gen{\`e}ve, Gen{\`e}ve,
Switzerland\\
\llap{$^b$}Astronomical Observatory, Jagiellonian University,
Krak{\'o}w, Poland\\
\llap{$^c$}Nicolaus Copernicus Astronomical Centre, Polish Academy of
  Sciences, Warsaw, Poland\\
\llap{$^d$}Astronomical Observatory, University of Warsaw, Warsaw,Poland\\
\llap{$^e$}Department of Information Technologies, Jagiellonian
University, Krak{\'o}w, Poland\\
\llap{$^f$}ISDC, Observatoire de Gen{\`e}ve, Universit{\'e} de
Gen{\`e}ve, Versoix, Switzerland\\
\llap{$^g$}Instytut Fizyki J{\c a}drowej im.~H.~Niewodnicza{\'n}skiego
  Polskiej Akademii Nauk, Krak{\'o}w, Poland\\
\llap{$^h$}Centrum Bada{\'n} Kosmicznych Polskiej Akademii Nauk,
Warsaw, Poland\\
\llap{$^i$}Astronomical Observatory, Taras Shevchenko National University of
Kyiv, Kyiv, Ukraine\\
\llap{$^k$}AGH University of Science and Technology, Krak{\'o}w, Poland\\
\llap{$^l$}Vrije Universiteit Brussels, Brussels, Belgium\\
\llap{$^m$}Institute of Physics of the Czech Academy of Sciences,
Prague, Czech Republic\\
\llap{$^n$}University College Dublin, Ireland\\
  e-mail: \email{moderski@camk.edu.pl}}

\abstract{The single mirror small-size telescope (SST-1M) is one of
  the telescope projects being proposed for the Cherenkov Telescope
  Array observatory by a sub-consortium of Polish and Swiss
  institutions. The SST-1M prototype structure is currently being
  constructed at the Institute of Nuclear Physics in Cracow, Poland,
  while the camera will be assembled at the University of Geneva,
  Switzerland. This prototype enables measurements of parameters
  having a decisive influence on the telescope performance. We present
  results of numerical simulations of the SST-1M performance based on
  such measurements. The telescope effective area, the expected
  trigger rates and the optical point spread function are calculated.}

\FullConference{The 34th International Cosmic Ray Conference\\
		30 July -- 6 August, 2015\\
		The Hague, The Netherlands}

\begin{document}

\section{Introduction}
The Cherenkov Telescope Array (CTA) observatory is an international
initiative to build the next generation of ground-based very high
energy gamma-ray instruments. Full sky coverage will be assured by two
arrays, with one site located in each of the northern and southern
hemispheres, and three main classes of telescopes (large, medium,
small) will cover the wide energy range from tens of GeV up to
hundreds of TeV \cite{ctaintro}.

The single mirror small-size telescope (SST-1M) is one of the
telescope projects being proposed for the CTA observatory by a
sub-consortium of Polish and Swiss institutions.  Its design is based
on the Davies-Cotton concept of multi-segment mirror composed of
spherical facets focused at twice the focal length of the telescope.
In case of the SST-1M telescope the mirror of $4\,$m diameter is
composed of $18$ hexagonal facets of $78\,$cm dimension
(flat-to-flat).  The focal length of the telescope is $5.6\,$m.

The prototype SST-1M telescope structure was installed at the
Institute of Nuclear Physics in Krak{\'o}w, Poland in November 2013
and is currently being tested and equipped with mirrors
\cite{nie15}. In parallel the SST-1M camera, named DigiCam, is being
constructed in the Universit{\'e} de Gen{\`e}ve, AGH University of
Science and Technology, and Astronomical Observatory of the
Jagiellonian University \cite{raj15}. The DigiCam utilizes silicon
photo-multipliers, which use in imaging atmospheric Cherenkov
telescopes was pioneered by the FACT telescope \cite{fact}, and fully
digital read-out and trigger electronics.

Based on measurements taken during the prototyping phase a series of
numerical simulations has been undertaken to determine the SST-1M
telescope performance.  During the simulations such parameters as
optical point spread function, night sky background trigger rate,
single telescope effective area and expected gamma-ray trigger rate
have been determined.

\section{Optical point spread function}
\begin{figure}[ht]
  \centering
  \begin{minipage}[t]{0.45\linewidth}
    \includegraphics[width=0.95\linewidth]{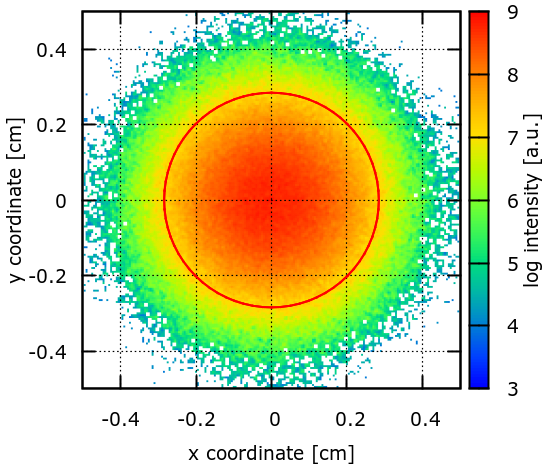}
    \caption{Simulated PSF shape for a single mirror facet -- real
      case corresponding to measured spot size $D_{80}(2f) = 5.7\,$mm
      marked with a red circle.}
    \label{fig:singlePSF}
  \end{minipage}
  \quad
  \begin{minipage}[t]{0.45\linewidth}
    \includegraphics[width=0.95\linewidth]{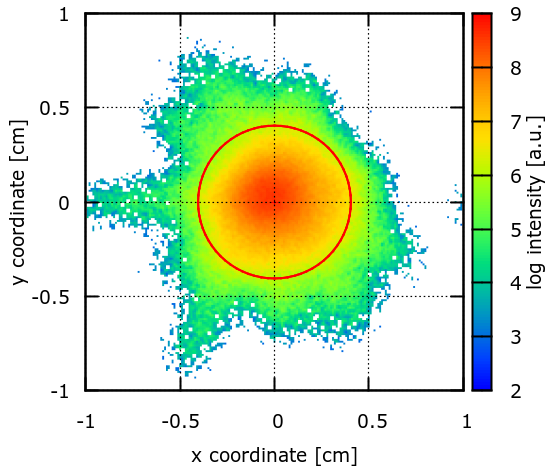}
    \caption{Example of simulated PSF shape for the whole SST-1M
      mirror. Red circle corresponds to $D_{80}(f) = 0.81\,$cm.}
    \label{fig:wholePSF}
  \end{minipage}
\end{figure}

The telescope optical point spread function (PSF) is a crucial
parameter to determine the ability to correctly image the air showers.
A ray tracing feature of the {\tt sim\_telarray} software package
\cite{ber08} has been used to determine the PSF of the SST-1M
telescope.  First a sample of $14$ mirror facets delivered for the
prototype has been measured in the Space Research Center in Warsaw to
provide initial values for the simulations.  An average spot size
$D_{80}$ at twice the focal length $f$ of $5.7 \pm 1.1\,$mm ($D_{80}$
being a diameter of the smallest circle containing $80\,$\% of the
light rays) has been measured.  This value has been used to fix the
simulation parameters.  The result of such a procedure is presented in
Fig.~\ref{fig:singlePSF}.  As a next step an additional measurement of
the average focal length of the mirror facets -- $<f> = 5.60 \pm
0.03\,$m has been used to simulate the PSF of the whole mirror.  A
value of $D_{80}(f) = 0.81 \pm 0.05\,$cm has been obtained, which
corresponds to the angular scale of $D_{80}(f) = 0.083 \pm 0.005\,$deg
for the on-axis rays.  An example of the PSF shape is presented in
Fig.~\ref{fig:wholePSF}.
    
\begin{figure}[ht]
  \centering
  \begin{minipage}[t]{0.45\linewidth}
    \includegraphics[width=0.95\linewidth]{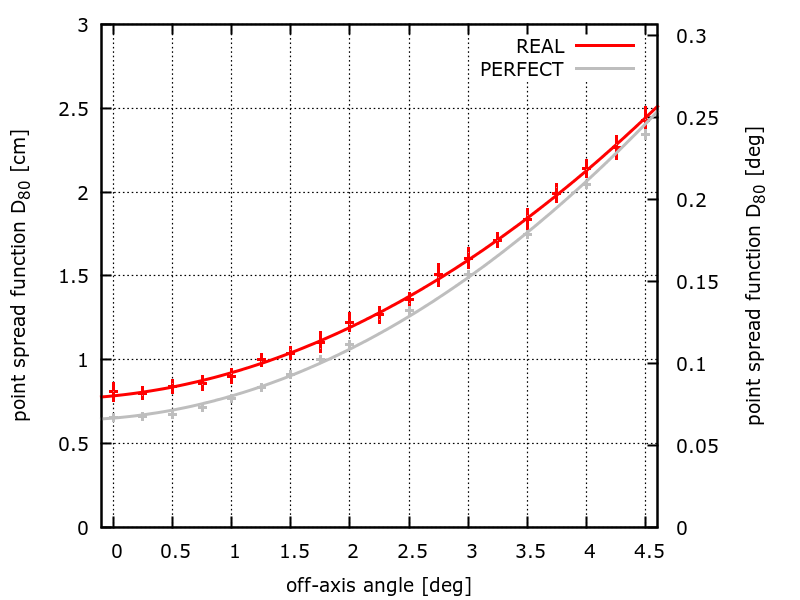}
    \caption{Comparison of PSF dependence on off-axis angle for both
      perfect and real cases.}
    \label{fig:psfoff}
  \end{minipage}
  \quad
  \begin{minipage}[t]{0.45\linewidth}
    \includegraphics[width=0.95\linewidth]{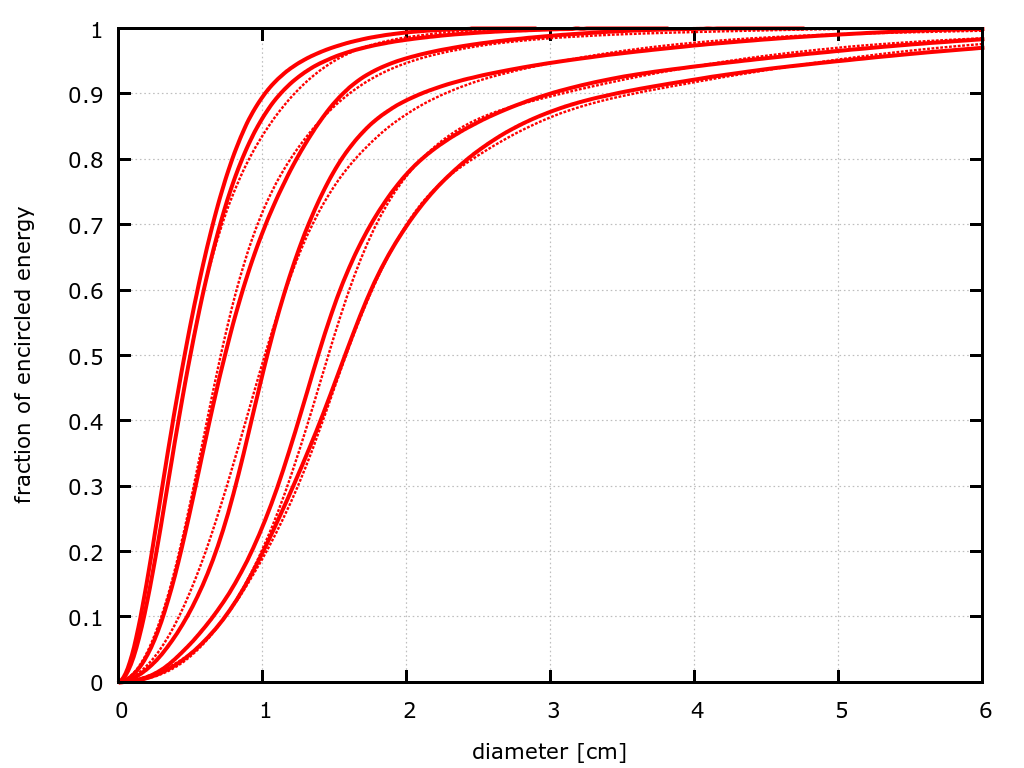}
    \caption{Fraction of encircled energy for off-axis rays at angles
      $0$, $1$, $2$, $3$, $4$, and $4.5\,$deg (from left to right,
      respectively).  Two curves for off-axis rays represent vertical
      (dotted lines) and horizontal (solid lines) axis.}
    \label{fig:encircled}
  \end{minipage}
\end{figure}

Additionally PSF values have been analysed for the off-axis rays in
the $80\,$\% of the camera field of view as required by the CTA
Consortium and compared to the case where all mirror facets are just
perfect.  The results are presented in Fig.~\ref{fig:psfoff}.  A
fraction of encircled energy has been also calculated as a function of
the off-axis angle and is presented in Fig.~\ref{fig:encircled}.

These studies show a very little PSF degradation as compared to the
ideal case and proof that the performance goals are quite easy
achievable with current mirror manufacturing techniques.

\section{Night sky background trigger rate}
A main source of night sky background (NSB) is the zodiacal light
being detected together with Cherenkov radiation from air showers.  In
the CTA framework the lowest flux of NSB light has been estimated to
be $0.24\,$ph$\,$ns$^{-1}\,$sr$^{-1}\,$cm$^{-2}$.  This flux level
corresponds to the so called ``dark night'' conditions, when
observations are performed without the presence of the moon
light. Taking into account the SST-1M telescope optical transmission,
effective collecting area of the mirror, angular size of the pixel,
and quantum efficiency of the photo detectors in the camera this value
corresponds to NSB level of $33.2 \times 10^6\,$ph$\,$s$^{-1}$ per
pixel.  This estimation does not take into acount additional component
coming from dark current counts.  Initial measurements show the level
of dark current of about $6\,$MHz, but the final estimation requires
additional studies.

\begin{figure}[ht]
  \centering
  \includegraphics[width=0.8\linewidth]{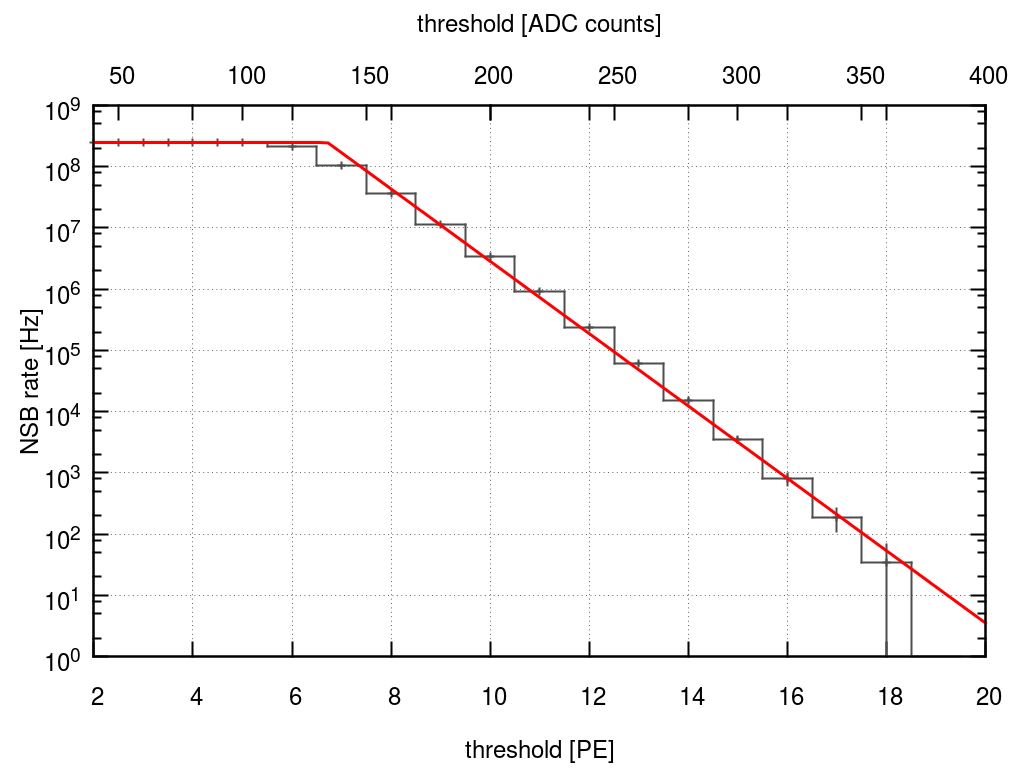}
  \caption{Night sky background trigger rate as a function of trigger
    threshold for SST-1M telescope.}
  \label{fig:nsbrate}
\end{figure}

Such conditions require efficient trigger system to lower the readout
rate and to avoid accidental coincidences.  Simulations of SST-1M
telescope detection process have been performed to determine the
dependence of the NSB trigger rate as a function of the trigger
threshold.  {\tt Sim\_telarray} package has been used and the trigger
topology has been defined in such a way that trigger conditions are
met if summed signal from a sector containing 7 neighboring pixels is
above the threshold.  Since DigiCam camera operates at $250\,$MHz
sampling rate this is also the highest NSB trigger rate which may be
expected from the system.  Due to constraints resulting from the
internal bandwidth limits this value needs to be lower to less than
about $20\,$kHz.  Results of the simulations are presented in
Fig.~\ref{fig:nsbrate} and in Tab.~\ref{tab:thlevels} we list trigger
thresholds required to lower the NSB rate to reasonable values.
\begin{table}[ht]
  \centering
  \includegraphics[width=0.8\linewidth]{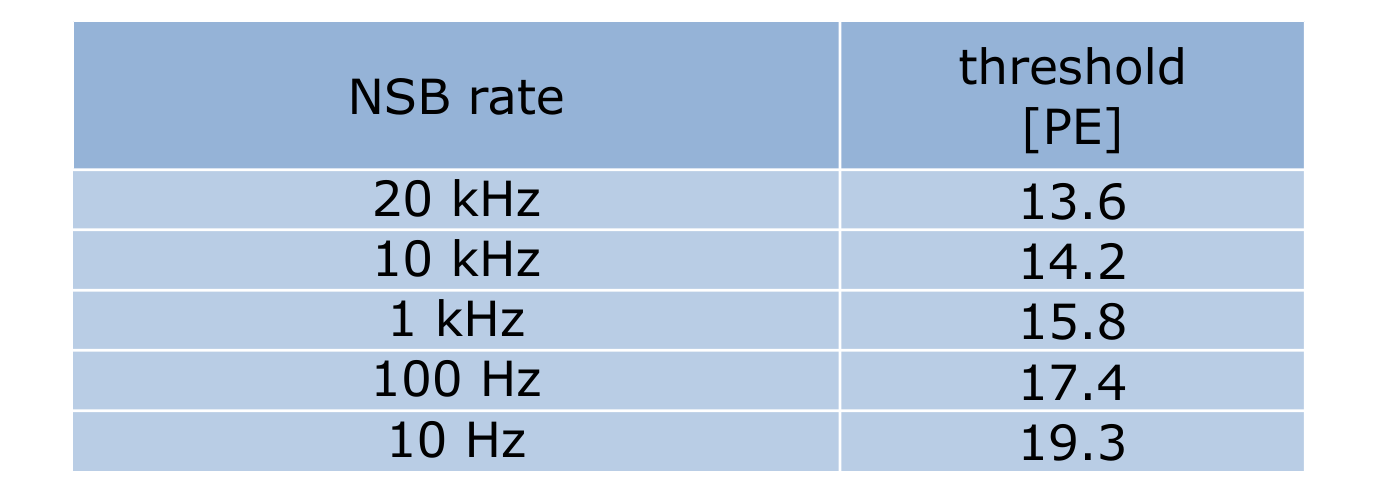}
  \caption{Trigger thresholds required to lower the trigger rate due to
  NSB.}
  \label{tab:thlevels}
\end{table}

During massive numerical simulations performed within Monte Carlo Work
Package of the CTA Consortium a concept of ``safe threshold'' has been
used \cite{ber13}. ``Safe threshold'' is defined as a threshold level
at which NSB trigger rate calculated for twice the NSB level equals
the cosmic ray trigger rate, which for that purpose is approximated as
$1.5$ times the proton trigger rate.  For the SST-1M telescope
prototype the ``safe threshold'' has been estimated equal to
$17.2\,$PE.

\section{Effective area, gamma rate and energy threshold}
\begin{figure}[ht]
  \centering
  \includegraphics[width=0.8\linewidth]{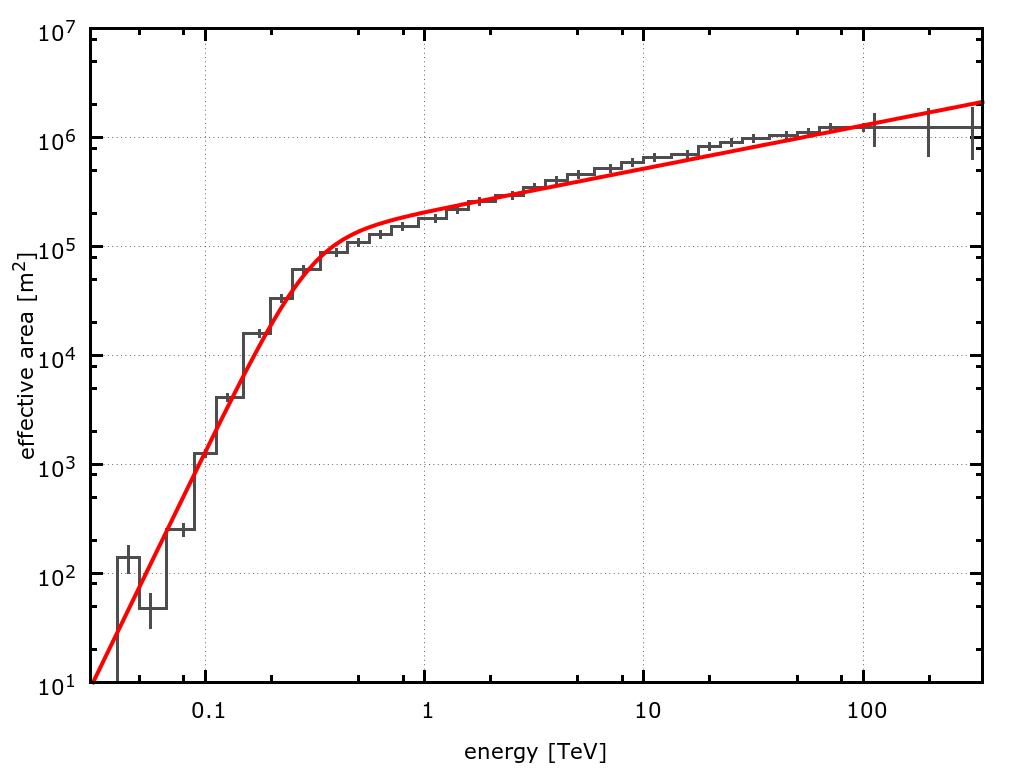}
  \caption{SST-1M telescope effective area as a function of energy.}
  \label{fig:effarea}
\end{figure}

Single telescope effective area for a point source of gamma-ray photons
is defined as
\begin{equation}
A_{{\rm eff},\gamma}(E) = 2 \pi \int_0^\infty P_\gamma(E,r) r {\rm d}
r \,,
\end{equation}
where $P_\gamma(E,r)$ is the detection probability (trigger
probability) for a gamma-ray shower induced by a primary photon with
energy $E$ and impact parameter $r$.  A total number of $1.6 \times
10^7$ gamma-ray showers within energy range $0.05-320\,$TeV has been
simulated with CORSIKA software \cite{corsika98} to estimate the
effective area of the SST-1M telescope. The result is presented in
Fig.~\ref{fig:effarea}.

\begin{figure}[ht]
  \centering
  \includegraphics[width=0.8\linewidth]{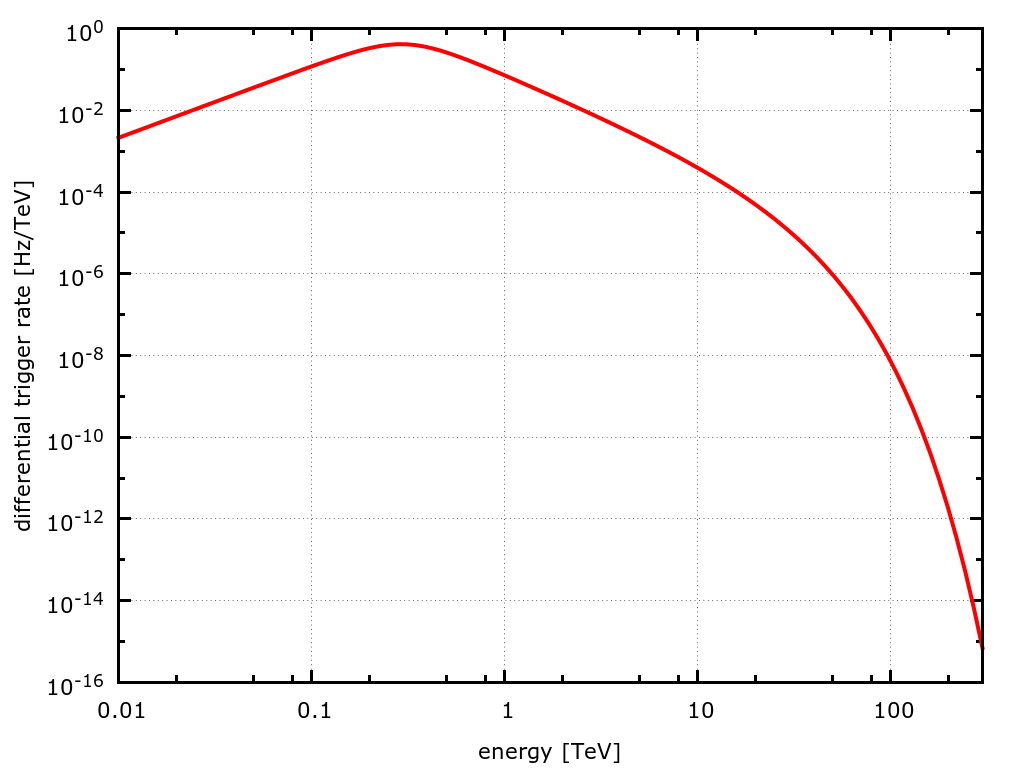}
  \caption{SST-1M telescope gamma-ray differential trigger rate for
    Crab-like point source as a function of energy.}
  \label{fig:diffrate}
\end{figure}

To estimate the expected gamma-ray trigger rate the above effective
area has been folded with a Crab-like source spectrum giving the
differential gamma rate plotted in Fig.~\ref{fig:diffrate}.  Thus the
total gamma-ray trigger rate, being the integral of the differential
trigger rate over all energies, for the SST-1M telescope is
$0.27\,$Hz.

Energy threshold, defined as the energy for which the differential
trigger rate reaches its maximum is $0.3\,$TeV.

\section{Summary}
SST-1M is a well advanced project.  Measurements of the system
components taken during the construction of the prototype allowed us
to perform a series of numerical simulations of the expected
performance of the telescope.  Results show that a network of SST-1M
telescopes is a valuable proposal for one of the sub-arrays of the
small size telescopes for the CTA Observatory.  In the near future
SST-1M parameters estimated during presented studies will be used in a
massive numerical simulations of the whole CTA array to determine the
final performance characteristics of the SST-1M sub-system.

\acknowledgments
We gratefully acknowledge support from the agencies and organizations
listed under Funding Agencies at this website:
http://www.cta-observatory.org/.  In particular we are grateful for
support from the Polish NCN grant DEC-2011/01/M/ST9/01891, the Polish
MNiSW grant no. 498/1/FNiTP/FNiTP/2010, the University of Geneva, and
the Swiss National Foundation.  This research was supported in part by
PLGrid Infrastructure.

\end{document}